\documentclass[conference,10pt]{IEEEtran}
\usepackage{epsfig,rotating,setspace,latexsym,amsmath,epsf,amssymb,amsfonts,bm,theorem,subfigure,epstopdf}
\usepackage{authblk,cite} 
\usepackage{bbm}
\usepackage{algorithmic,algorithm}
\usepackage{color}

\IEEEoverridecommandlockouts
\allowdisplaybreaks

\begin{document}
\bstctlcite{IEEEexample:BSTcontrol}

	%
	\title{Over-the-Air Adversarial Attacks on Deep Learning Based Modulation Classifier over Wireless Channels}

	
	\author[1]{Brian Kim}
	\author[2]{Yalin E. Sagduyu}
	\author[2]{Kemal Davaslioglu}
	\author[2]{Tugba Erpek}
	\author[1]{Sennur Ulukus}
	\affil[1]{\normalsize Department of Electrical and Computer Engineering, University of Maryland, College Park, MD 20742, USA}
	\affil[2]{\normalsize Intelligent Automation, Inc., Rockville, MD 20855, USA \thanks{This effort is supported by the U.S. Army Research Office under contract W911NF-17-C-0090. The content of the information does not necessarily reflect the position or the policy of the U.S. Government, and no official endorsement should be inferred.}}
	
	\maketitle


\begin{abstract}
	We consider a wireless communication system that consists of a transmitter, a receiver, and an adversary. The transmitter transmits signals with different modulation types, while the receiver classifies its received signals to modulation types using a deep learning-based classifier. In the meantime, the adversary makes over-the-air transmissions that are received as superimposed with the transmitter's signals to fool the classifier at the receiver into making errors. While this evasion attack has received growing interest recently, the channel effects from the adversary to the receiver have been ignored so far such that the previous attack mechanisms cannot be applied under realistic channel effects. In this paper, we present how to launch a realistic evasion attack by considering channels from the adversary to the receiver. Our results show that modulation classification is vulnerable to an adversarial attack over a wireless channel that is modeled as Rayleigh fading with path loss and shadowing. We present various adversarial attacks with respect to availability of information about channel, transmitter input, and classifier architecture. First, we present two types of adversarial attacks, namely a targeted attack (with minimum power) and non-targeted attack that aims to change the classification to a target label or to any other label other than the true label, respectively. Both are white-box attacks that are transmitter input-specific and use channel information. Then we introduce an algorithm to generate adversarial attacks using limited channel information where the adversary only knows the channel distribution. Finally, we present a black-box universal adversarial perturbation (UAP) attack where the adversary has limited knowledge about both channel and transmitter input. By accounting for different levels of information availability, we show the vulnerability of modulation classifier to over-the-air adversarial attacks.  
\end{abstract}

\section{Introduction}
Advances in \emph{deep learning} (DL) based on \emph{deep neural networks} (DNNs) have supported numerous applications to learn from complex data domains such as in computer vision and speech recognition 
\cite{Goodfellow1}. Following the success of these applications, DL has been applied to wireless communications, where channel, interference, and traffic effects jointly contribute to the high complexity of the spectrum data \cite{erpek1}.


Machine learning (ML) in the presence of adversaries have been studied in the context of \emph{adversarial machine learning} \cite{Vorobeychik1}. In particular, DNNs are known to be highly susceptible to adversarial attacks, as demonstrated with applications in computer vision domain \cite{Szegedy1}. 

Recently, adversarial ML has been studied in wireless communication systems using DNNs. \emph{Exploratory (inference) attacks} have been considered in \cite{Terpek}, where an adversary builds a DNN to learn the transmission pattern in the channel and jam transmissions that would be otherwise successful.   Over-the-air spectrum \emph{poisoning (causative) attacks} have been considered in \cite{Sagduyu1}, where an adversary poisons (falsifies) a transmitter's spectrum sensing data over the air by transmitting during the short spectrum sensing period of the transmitter. \emph{Trojan attacks} have been studied in \cite{Davaslioglu1} against a signal classifier, where an adversary slightly manipulates training data by inserting Trojans in terms of modifying the phases and the labels of only few training data to a target label, and then transmits signals with the same phase shift in the inference time to fool the signal classifier. 

Built upon adversarial ML, \emph{adversarial attacks} (a.k.a \emph{evasion attacks}) correspond to small modifications of the original input to the DNNs that make DL algorithm to misclassify the input. These small modifications are not just random noise but carefully designed in a way that changes the decision of the DL algorithm.
As an evasion attack, \cite{Larsson1} has showed that the end-to-end autoencoder communication systems, proposed in \cite{Oshea1}, are vulnerable to adversarial attacks in an additive white Gaussian noise (AWGN) channel environment, where the attack increases the block-error-rate at the receiver. Also, adversarial attacks have been studied for modulation classification of wireless signals in \cite{Larsson2}, where fast gradient method (FGM) \cite{Kurakin1} is used to generate adversarial attacks. Specifically, targeted FGM attack has been used by enforcing the DNNs to misclassify the input signals to a target label. Here, the target is decided by searching over all possible target labels and selecting the one with the least perturbation required to enforce misclassification. It has been shown that the modulation classifier used in \cite{Oshea2} incurs major errors due to adversarial attacks in the AWGN channel. 

Similar evasion attacks and corresponding defense mechanisms have been studied in \cite{Flowers1,Bair1,Kokalj1,Kokalj2,Kokalj3,Gunduz1}. Previous work has considered the AWGN channel from the transmitter to the receiver only, but has not considered channel effects (path loss or fading) from the adversary to the receiver. However, even a small channel effect would significantly reduce the impact of adversarial attacks by reducing the received perturbation power just below the necessary level such that the adversarial attack fails in changing classification decision over the air.  


In this paper, we consider a wireless communication system where a DNN is used to classify wireless signals to modulation types as in \cite{Larsson2}, and show how to make this classifier vulnerable to adversarial attacks even in the presence of \emph{realistic channel effects} from the adversary to the receiver. For that purpose, we design adversarial attacks with a power constraint that decreases the accuracy of detecting modulation type at the receiver. We first propose two white-box attacks, a \emph{targeted attack} with minimum power and a \emph{non-targeted attack}, subject to channel effects known by the adversary. We show that the adversarial attack fails if the channel between the adversary and the receiver is not considered (as in \cite{Larsson2, Flowers1,Bair1,Kokalj1,Kokalj2,Kokalj3}) when designing the adversarial attack. Then we show how to design the adversarial attack by accounting for known channel effects.

Next, we relax the assumption that the adversary knows the exact channel condition and present a white-box adversarial attack with \emph{limited channel information} available at the adversary. Finally, we design a \emph{black-box universal adversarial perturbation (UAP) attack}, where the adversary has limited information about the transmitter input, channel conditions, and the classifier architecture. All these attacks demonstrate the importance of channel effects on attack performance and raise the need to utilize channel information in designing adversarial attacks and launching them over the air. 

The rest of the paper is organized as follows. Section \ref{sec:SystemModel} explains the system model. Sections \ref{sec:TargetedAttackWChInfo} and \ref{sec:NontargetedAttackWChInfo} describe targeted and non-targeted white-box adversarial attacks, respectively, using channel information. Section \ref{sec:WhiteBoxWOutChInfo} considers the white-box adversarial attack with limited channel information. Section \ref{sec:UniversalAttack} describes a universal adversarial perturbation. Section \ref{sec:SimResults} presents simulation results. Section \ref{sec:Conclusion} concludes the paper.   


\section{System Model} \label{sec:SystemModel}
We consider a wireless communication system that consists of a transmitter, a receiver, and an adversary. The transmitter transmits signals with one of the modulation types. The receiver applies a pre-trained DL-based classifier on the received signals to classify the modulation type that is used at the transmitter. The adversary launches an attack by transmitting over the air to cause misclassification at the receiver. 

The DNN classifier at the receiver is denoted by $f(\cdot;\boldsymbol{\theta}): \mathcal{X} \rightarrow \mathbb{R}^{C}$, where $\boldsymbol{\theta}$ is the parameters of the DNN and $C$ is the number of modulation types. Note $\mathcal{X} \subset \mathbb{C}^{p}$, where $p$ is the dimension of the complex-valued (in-phase/quadrature) inputs that can be also represented by concatenation of two real-valued inputs. The classifier $f$ assigns a modulation type $\hat{l}(\boldsymbol{x},\boldsymbol{\theta}) = \arg \max_{k} f_{k}(\boldsymbol{x},\boldsymbol{\theta})$ to every input $\boldsymbol{x}\in \mathcal{X}$. In this formulation, $f_{k}(\boldsymbol{x},\boldsymbol{\theta})$ is the output of classifier $f$ corresponding to the $k$th modulation type.

There exist the channel $\boldsymbol{h}_{tr}$ from the transmitter to the receiver and the channel $\boldsymbol{h}_{ar}$ from the adversary to the receiver, where $\boldsymbol{h}_{tr} = [h_{tr,1}, h_{tr,2},\cdot \cdot \cdot,h_{tr,p}]^{T}\in \mathbb{C}^{p\times 1}$ and $\boldsymbol{h}_{ar} = [h_{ar,1}, h_{ar,2},\cdot \cdot \cdot,h_{ar,p}]^{T}\in \mathbb{C}^{p\times 1}$. If the transmitter transmits $\boldsymbol{x}$, the receiver receives $\boldsymbol{r}_{t} = \mathbf{H}_{tr} \boldsymbol{x}+\boldsymbol{n}$, if there is no adversarial attack, or  receives $\boldsymbol{r}_{a} = \mathbf{H}_{tr} \boldsymbol{x}+ \mathbf{H}_{ar}\boldsymbol{\delta}+\boldsymbol{n}$, if the transmitter launches an adversarial attack by transmitting the perturbation signal $\boldsymbol{\delta}$, where $\mathbf{H}_{tr} = \mbox{diag} \{h_{tr,1},\cdot\cdot\cdot, h_{tr,p}\}\in \mathbb{C}^{p\times p}, \mathbf{H}_{ar} = \mbox{diag}\{h_{ar,1},\cdot \cdot\cdot,h_{ar,p}\}\in \mathbb{C}^{p \times p}$, $\boldsymbol{\delta}\in \mathbb{C}^{p\times1}$ and $\boldsymbol{n}\in \mathbb{C}^{p\times 1}$ is complex Gaussian noise. For a stealth  (i.e., difficult to detect) attack, the adversarial perturbation $\boldsymbol{\delta}$ is restricted as $||\boldsymbol{\delta}||^{2}_{2}\le P_{\textit{max}}$ for some suitable power $P_{\textit{max}}$.  
The adversary obtains the adversarial perturbation $\boldsymbol{\delta}$ for the input $\boldsymbol{x}$ and classifier $f$ by solving the following optimization problem:
\begin{align}
\max_{\delta}& \quad \mathbb{I}\{\hat{l}(\boldsymbol{r}_{t},\boldsymbol{\theta}) \ne \hat{l}(\boldsymbol{r}_{a},\boldsymbol{\theta})  \} \nonumber\\
\mbox{subject to} & \quad ||\boldsymbol{\delta}||^{2}_{2} \le P_{\textit{max}},
\end{align}
where $\mathbb{I}\{\cdot\}$ is an indicator function.


In practice solving (1) is difficult. Thus, different methods have been proposed (primarily in the computer vision domain) to approximate the adversarial perturbation such as FGM. FGM is a computationally efficient method for crafting adversarial attacks by linearizing the loss function of the DNN classifier. Let $L(\boldsymbol{\theta},\boldsymbol{x},\boldsymbol{y})$  denote the loss function of the model, where $\boldsymbol{y}\in \{0,1 \}^C$ is the label vector. Then FGM linearizes the loss function in a neighborhood of $\boldsymbol{x}$ and uses this linearized function for optimization.

There are two types of attacks called \emph{targeted attacks} and \emph{non-targeted attacks} that involve different objective functions to optimize. In a targeted attack, the adversary is trying to generate a perturbation that causes the classifier at the receiver to have a specific misclassification, e.g., the classifier classifies QPSK modulation as QAM16, whereas in non-targeted FGM attack, the adversary is searching for a perturbation that causes any misclassification (independent of target label). We will further explain these two types of attacks in the next section.

Our  goal  in  this  paper  is  to  design  an  attack  to  fool  the classifier at the receiver while considering the channel effects and satisfying the power constraint at  the  adversary. For the white-box adversarial attacks, we assume that the adversary knows the architecture ($\boldsymbol{\theta}$ and $L(\cdot)$) of the classifier at the receiver. Also, we assume that the adversary knows the input at the receiver and consequently the channel $\boldsymbol{h}_{ar}$ between the adversary and the receiver. We will relax these assumptions in the later part of the paper. 

\section{Targeted White-Box Adversarial Attacks using Channel Information} \label{sec:TargetedAttackWChInfo}
For the targeted attack, the adversary minimizes $L(\boldsymbol{\theta},\boldsymbol{r}_{a},\boldsymbol{y}^{\textit{target}})$ with respect to $\boldsymbol{\delta}$, where $\boldsymbol{y}^{\textit{target}}$ is one-hot encoded desired target class. FGM is used to linearize the loss function as $L(\boldsymbol{\theta},\boldsymbol{r}_{a},\boldsymbol{y}^{\textit{target}}) \approx  L(\boldsymbol{\theta},\boldsymbol{r}_{t},\boldsymbol{y}^{\textit{target}}) + (\mathbf{H}_{ar}\boldsymbol{\delta})^{T} \nabla_{\boldsymbol{x}}L(\boldsymbol{\theta},\boldsymbol{r}_{t},\boldsymbol{y}^{\textit{target}}) $ that is minimized by setting $\mathbf{H}_{ar}\boldsymbol{\delta} = -\alpha \nabla_{\boldsymbol{x}}L(\boldsymbol{\theta},\boldsymbol{r}_{t},\boldsymbol{y}^{\textit{target}})$, where $\alpha$ is a scaling factor to constrain the adversarial perturbation power to $P_{\textit{max}}$. 

The adversary can generate different targeted attacks with respect to different $\boldsymbol{y}^{\textit{target}}$ that causes the classifier at the receiver to misclassify the received signals to $C-1$ different modulation types. Thus, as in \cite{Larsson2}, the adversary can create targeted attacks for all $C-1$ modulation types and chooses the target modulation that uses the least power. However, \cite{Larsson2} only considered the AWGN channel, i.e., $\mathbf{H}_{ar} = \mathbf{I}$, which falls short from representing channel conditions encountered in real wireless communication systems. We call the targeted attack perturbation in \cite{Larsson2} as $\boldsymbol{\delta}^{\textit{NoCh}}$, which is an optimal targeted attack under the AWGN channel. The detailed algorithm is given in Algorithm 1 by setting $\mathbf{H}_{ar} = \mathbf{I}$. In the following subsections, we propose three targeted adversarial attacks to overcome the random effects of the channel. 
\subsection{Channel Inversion Attack}
We first begin with a naive attack, where the adversary designs its attack by inverting the channel in the optimal targeted attack $\boldsymbol{\delta}^{\textit{NoCh}}$, which is obtained using Algorithm 1 with the AWGN channel. Since the adversarial attack goes through channel $\boldsymbol{h}_{ar}$,  the $i$th element of the perturbation $\boldsymbol{\delta}$ is simply designed as ${\delta}_{i} = \frac{{\delta}_{i}^{\textit{NoCh}}}{{h}_{ar,i}}$  so that after going through the channel it has the same direction as ${\delta}^{\textit{NoCh}}_{i}$ for $i = 1,\cdot\cdot\cdot, p$. Furthermore, in order to satisfy the power constraint $P_{\textit{max}}$, we introduce a scaling factor $\alpha$ so that $\boldsymbol{\delta}^{div} = -\alpha \boldsymbol{\delta}$, where $\alpha = \frac{\sqrt{P_{\textit{max}}}}{||\boldsymbol{\delta}||_{2}}$ to satisfy the power constraint at the adversary. Thus, the attack received at the receiver is  $\mathbf{H}_{ar}\boldsymbol{\delta}^{div} = -\alpha \boldsymbol{\delta}^{\textit{NoCh}}$.

%

\subsection{Minimum Mean Squared Error (MMSE) Attack}
In the MMSE attack, the adversary designs the perturbation $\boldsymbol{\delta}^{\textit{\textit{MMSE}}}$ so that the distance between the attack after going through the channel and the optimal targeted attack over AWGN channel is minimized. By designing the attack in this way, the received attack at the receiver is close to the optimal targeted attack as much as possible while satisfying the power constraint at the adversary. However, since the classifier is sensitive to not only the direction but also the power of perturbation, the squared error criterion might penalize the candidates of $\boldsymbol{\delta}^{\textit{MMSE}}$, which have more power with the direction of $\boldsymbol{\delta}^{\textit{NoCh}}$, i.e., $\boldsymbol{\delta}^{\textit{MMSE}} = \gamma\boldsymbol{\delta}^{\textit{NoCh}}$. Therefore, we formulate the optimization problem to select the perturbation $\boldsymbol{\delta}^{\textit{MMSE}}$ as
\begin{align}
 \min_{\boldsymbol{\delta}^{\textit{MMSE}}}&  \quad || \mathbf{H}_{ar}\boldsymbol{\delta}^{\textit{MMSE}} - \gamma\boldsymbol{\delta}^{\textit{NoCh}}||^{2}_{2}\nonumber\\
 \mbox{subject to}& \quad \quad ||\boldsymbol{\delta}^{\textit{MMSE}} ||^{2}_{2} \le P_{\textit{max}},
\end{align}
where $\gamma$ is optimized by line search. We can write (2) as
	\begin{align}
	 \min_{{\delta}_{i}^{\textit{MMSE}}} & \quad  \sum^{p}_{i=1} ||h_{ar,i}\delta^{\textit{MMSE}}_{i}-\gamma\delta^{\textit{NoCh}}_{i}||_2^2             \nonumber\\
	 \mbox{subject to} & \quad \quad  \sum^{p}_{i=1}||\delta^{\textit{MMSE}}_{i}||_2^{2}  \le P_{\textit{max}}     .     
    \end{align}
 We solve the convex optimization problem (3) by using Lagrangian method. The Lagrangian for (3) is given by
	\begin{align}
	\mathcal{L}= & \sum^{p}_{i=1} ||h_{ar,i}\delta^{\textit{MMSE}}_{i}\hspace{-0.5mm}-\hspace{-0.5mm}\gamma\delta^{\textit{NoCh}}_{i}||_2^2 \hspace{-0.5mm}+\hspace{-0.5mm} \lambda \left( \sum^{p}_{i=1}||\delta^{\textit{MMSE}}_{i}||_2^{2}\hspace{-0.8mm} -\hspace{-0.8mm} P_{\textit{max}}\right),
	\end{align}
	where $\lambda\ge 0$. The KKT conditions are given by
	\begin{equation}
	 h^{*}_{ar,i}(h_{ar,i}\delta_{i}^{\textit{MMSE}} - \gamma\delta_{i}^{\textit{NoCh}})+\lambda\delta_{i}^{\textit{MMSE}} = 0 ,
	\end{equation}  
	for all $i = 1,\cdot\cdot\cdot, p$. From KKT conditions, we obtain the perturbation of the MMSE attack as

	\begin{equation}
	\delta_{i}^{\textit{MMSE}} = -\frac{\gamma h^{*}_{ar,i}\delta_{i}^{\textit{NoCh}}}{h^{*}_{ar,i}h_{ar,i}+ \lambda},
	\end{equation} 
	for all $i = 1,\cdot\cdot\cdot, p$, where $\lambda$ is determined by the power constraint at the adversary. Note that the received perturbation at the receiver is $\mathbf{H}_{ar}\boldsymbol{\delta}^{\textit{MMSE}} = -\boldsymbol{\alpha}^{T} \boldsymbol{\delta}^{\textit{NoCh}}$ where $\boldsymbol{\alpha} \in \mathbb{R}^{p\times 1}$ and each element of $\boldsymbol{\alpha}$ is $\alpha_{i} = \frac{\gamma h_{ar,i}h^{*}_{ar,i}}{h^{*}_{ar,i}h_{ar,i}+ \lambda} $. 


\begin{algorithm}[t]
	\setstretch{1.15}
	\caption{MRPP attack}
	\label{target}
	\begin{algorithmic}[1] 
	    \STATE Inputs: input $\boldsymbol{r}_{t}$, desired accuracy $\varepsilon_{acc}$, power constraint $P_{\textit{max}}$ and model of the classifier
		\STATE Initialize: $ \boldsymbol{\varepsilon}\leftarrow \boldsymbol{0}^{C \times 1}$
		\FOR{class-index $c$ in range($C$)} 
		\STATE $\varepsilon_{\textit{max}} \leftarrow P_{\textit{max}}, \varepsilon_{min} \leftarrow 0$
		\STATE $\boldsymbol{\delta}_{\textit{norm}}^{c} =\frac{\mathbf{H}^{*}_{ar}\nabla_{\boldsymbol{x}}L(\boldsymbol{\theta},\boldsymbol{r}_{t},\boldsymbol{y}^{c})}{(||\mathbf{H}^{*}_{ar}\nabla_{\boldsymbol{x}}L(\boldsymbol{\theta},\boldsymbol{r}_{t},\boldsymbol{y}^{c})||_{2})}$
		\WHILE{$\varepsilon_{\textit{max}}-\varepsilon_{min} > \varepsilon_{acc}$}
		\STATE $\varepsilon_{avg} \leftarrow (\varepsilon_{\textit{max}}+\varepsilon_{min})/2$
		\STATE $\boldsymbol{x}_{adv} \leftarrow \boldsymbol{x} - \varepsilon_{avg}\mathbf{H}_{ar}\boldsymbol{\delta}_{\textit{norm}}^{c}$
		\IF{$\hat{l}(\boldsymbol{x}_{adv})== l_{\textit{true}}$}
		\STATE $\varepsilon_{min}\leftarrow \varepsilon_{avg}$
		\ELSE
		\STATE $\varepsilon_{\textit{max}}\leftarrow \varepsilon_{avg}$
		\ENDIF
		\ENDWHILE
		\STATE $\varepsilon[c] = \varepsilon_{\textit{max}}$
		\ENDFOR 
		\STATE $\textit{target} = \arg\min \boldsymbol{\varepsilon}, \;\boldsymbol{\delta}^{\textit{MRPP}} = -\sqrt{P_{\textit{max}}}\boldsymbol{\delta}_{\textit{norm}}^{\textit{target}}$ 
		
	\end{algorithmic}
\end{algorithm}

\subsection{Maximum Received Perturbation Power (MRPP) Attack}
In the MRPP attack, the adversary selects the perturbation $\boldsymbol{\delta}^{\textit{MRPP}}$ to maximize the received perturbation power at the receiver and analyzes how the received perturbation power affects the decision process of the classifier. 
To maximize the received perturbation power and effectively fool the classifier into making a specific classification error, the adversary has to fully utilize the channel between the adversary and the receiver. Thus, if the targeted attack ${\delta}^{\textit{target}}_{i}$ is multiplied by the conjugate of the channel, ${h}^{*}_{ar,i}$, then the received perturbation after going through the channel becomes $||{h}_{ar,i}||^{2}_{2}{\delta}^{\textit{target}}_{i}$. In this attack, not only the direction is unaffected after going through the channel but also the power is maximized by utilizing the channel. Finally, the adversary generates targeted attack for every possible modulation type to decide the target class and calculate the scaling factor to satisfy the power constraint at the adversary. The details are presented in Algorithm 1.


\section{Non-Targeted White-Box Adversarial Attacks using Channel Information} \label{sec:NontargetedAttackWChInfo}
In this section, the adversary designs the attack based on the non-targeted attack and its objective is to maximize the loss function $L(\boldsymbol{\theta},\boldsymbol{r}_{a},\boldsymbol{y}^{\textit{true}})$, where $\boldsymbol{y}^{\textit{true}}$ is the true label of $\boldsymbol{x}$. FGM is used to linearize the loss function as $L(\boldsymbol{\theta},\boldsymbol{r}_{a},\boldsymbol{y}^{\textit{true}}) \approx L(\boldsymbol{\theta},\boldsymbol{r}_{t},\boldsymbol{y}^{\textit{true}}) + (\mathbf{H}_{ar}\boldsymbol{\delta})^{T} \nabla_{\boldsymbol{x}}L(\boldsymbol{\theta},\boldsymbol{r}_{t},\boldsymbol{y}^{\textit{true}})$ that is maximized by setting $\mathbf{H}_{ar}\boldsymbol{\delta} = \alpha \nabla_{\boldsymbol{x}}L(\boldsymbol{\theta},\boldsymbol{r}_{t},\boldsymbol{y}^{\textit{true}})$, where $\alpha$ is a scaling factor to constrain the adversarial perturbation power to $P_{\textit{max}}$. Based on FGM for the non-targeted attack, we propose non-targeted adversarial attacks to effectively attack the classifier at the receiver.

\begin{algorithm}[t]
	\setstretch{1.15}
	\caption{Crafting naive non-targeted attack}
	\label{non_target}
	\begin{algorithmic}[1]  
		\STATE Inputs: number of epochs $E$, power constraint $P_{\textit{max}}$, true label $\boldsymbol{y}^{\textit{true}}$ and model of the classifier
		\STATE Initialize: Sum of gradient $\Delta \leftarrow 0 $ , $\boldsymbol{x}\leftarrow \boldsymbol{r}_{t}$
		\FOR{epoch $e$ in range($E$)} 
		\STATE $\boldsymbol{\delta}_{\textit{norm}} =\frac{\nabla_{\boldsymbol{x}}L(\boldsymbol{\theta},\boldsymbol{x},\boldsymbol{y}^{\textit{true}})}{(||\nabla_{\boldsymbol{x}}L(\boldsymbol{\theta},\boldsymbol{x},\boldsymbol{y}^{\textit{true}})||_{2})}$
		\STATE $\boldsymbol{x} \leftarrow \boldsymbol{x} + \sqrt{\frac{{P_{\textit{max}}}}{E}}\mathbf{H}_{ar}\boldsymbol{\delta}_{\textit{norm}}$
		\STATE $\Delta \leftarrow \Delta+ \sqrt{\frac{{P_{\textit{max}}}}{E}}\boldsymbol{\delta}_{\textit{norm}}$
		\ENDFOR 
		\STATE $\boldsymbol{\delta}^{\textit{naive}} = \sqrt{P_{\textit{max}}}\frac{\boldsymbol{\Delta}}{||\boldsymbol{\Delta}||_{2}} $
		
	\end{algorithmic}
\end{algorithm}

\subsection{Naive Non-Targeted Attack}
As in the targeted attacks, we begin with the naive non-targeted attack. First, the adversary divides its power $P_{\textit{max}}$ into $E$ epochs and uses $\frac{{P_{\textit{max}}}}{E}$ amount of power to the gradient of loss function to tilt the transmitted signal from the transmitter. Next, the adversary calculates the gradient again with respect to the transmitted signal from the transmitter and added perturbation. Then the adversary adds another perturbation with power $\frac{{P_{\textit{max}}}}{E}$ using the new gradient. This scheme generates the best direction to increase the loss function at that specific instance. Finally, the adversary repeats this procedure $E$ times and sums all the gradients of the loss function that were added to the transmitted signal from the transmitter since the adversary can send only one perturbation at a time over the air. Finally, a scaling factor is introduced to satisfy the power constraint at the adversary. The details of this algorithm are presented in Algorithm \ref{non_target}.

\subsection{Minimum Mean Squared Error (MMSE) Attack}
The non-targeted MMSE attack is designed similar to the targeted MMSE attack. The adversary first obtains $\boldsymbol{\delta}^{\textit{NoCh}}$ from the naive non-targeted attack with $\mathbf{H}_{ar} = \mathbf{I}$ and uses it to solve problem (2). Thus, the solution is the same as the solution to (2) except that it has the opposite direction to maximize the loss function, whereas the loss function is minimized for the targeted attack case. Therefore, the perturbation selected by the MMSE scheme for non-targeted attack is $\boldsymbol{\delta}^{\textit{MMSE}} =  \boldsymbol{\alpha}^{T} \boldsymbol{\delta}^{\textit{NoCh}}$, where $\boldsymbol{\alpha} \in \mathbb{R}^{p}$ and each element of $\boldsymbol{\alpha}$ is $\alpha_{i} = \frac{\gamma h^{*}_{ar,i}}{h^{*}_{ar,i}h_{ar,i}+ \lambda}$.

\subsection{Maximum Received Perturbation Power (MRPP) Attack}
As we have seen in the targeted MRPP attack, the attack should be in the form of $\boldsymbol{\delta}^{\textit{MRPP}} = \mathbf{H}^{*}_{ar}\boldsymbol{\delta}^{target}$ to maximize the received perturbation power at the receiver. Thus, the naive non-targeted attack is changed to create the MRPP non-targeted attack by changing $\boldsymbol{\delta}_{\textit{norm}}$ in Algorithm \ref{non_target} to $\frac{\mathbf{H}_{ar}^{*}\nabla_{\boldsymbol{x}}L(\boldsymbol{\theta},\boldsymbol{x},\boldsymbol{y}^{\textit{true}})}{(||\mathbf{H}_{ar}^{*}\nabla_{\boldsymbol{x}}L(\boldsymbol{\theta},\boldsymbol{x},\boldsymbol{y}^{\textit{true}})||_{2})}$.

\section{White-box Adversarial Attack with Limited Channel Information} \label{sec:WhiteBoxWOutChInfo}
The adversarial attacks that are designed in the previous sections use the exact channel information. However, this may not always be the case in practical scenarios. Therefore, in this section, we propose an algorithm to generate adversarial attacks using principal component analysis (PCA) with limited channel information, i.e., distribution of the channel. PCA was also used in \cite{Larsson2} for the AWGN channel case only. PCA is performed by eigenvalue decomposition of the data covariance matrix or singular value decomposition of a data matrix and is used to obtain the principal component which has the largest variance. In other words, PCA finds the principal component that provides the most information about the data with reduced dimension by projecting the data onto it. 

To generate an adversarial attack with limited channel information, we first generate $N$ realizations of the channel between the adversary and the receiver $\{\mathbf{H}_{ar}^{(1)}, \mathbf{H}_{ar}^{(2)}, \cdot\cdot\cdot, \mathbf{H}_{ar}^{(N)}\}$ from a known distribution. Then we generate $N$ adversarial attacks using white-box attack algorithms from the previous sections, either targeted or non-targeted, using $N$ realizations of the channel and the known input at the classifier $\boldsymbol{r}_t$. Finally, we stack $N$ generated adversarial attacks in a matrix and find the principal component of the matrix to use it as the adversarial attack with limited channel information. The details are presented in Algorithm \ref{limitch}.

\begin{algorithm}[t]
	\setstretch{1.15}
	\caption{Crafting adversarial attack with limited channel information}
	\label{limitch}
	\begin{algorithmic}[1]  
		\STATE Inputs:  $N$ channel realization $\{\mathbf{H}_{ar}^{(1)}, \cdot\cdot\cdot, \mathbf{H}_{ar}^{(N)}\}$, input $\boldsymbol{r}_{t}$ and model of the classifier
		\STATE Initialize: $\Delta \leftarrow 0 $
		\FOR {$n$ in range($N$)} 
		\STATE Find $\boldsymbol{\delta}^{(n)}$ from white-box attack algorithm using $\boldsymbol{r}_{t}$ and $\mathbf{H}_{ar}^{(n)}$
		\STATE Stack $\boldsymbol{\delta}^{(n)}$ to $\Delta$
		\ENDFOR 
		\STATE Compute the first principle direction $\boldsymbol{v}_{1}$ of $\Delta$ using PCA 
		\STATE $\Delta = \mathbf{U}\mathbf{\Sigma}\mathbf{V}^{T}$ and $\boldsymbol{v}_{1} = \mathbf{V}\boldsymbol{e}_{1}$
		\STATE $\boldsymbol{\delta}^{\textit{limited}} = \sqrt{P_{\textit{max}}}\boldsymbol{v}_{1}$
	\end{algorithmic}
\end{algorithm}

\section{Universal Adversarial Attack} \label{sec:UniversalAttack}
In the previous sections, the adversary designs a white-box attack with the assumptions that it knows the architecture of the classifier at the receiver, the channel between the adversary and the receiver, and the exact input at the receiver. However, these assumptions are not always practical in real wireless communications systems. Thus, in this section, we relax these assumptions and propose UAP attacks.
\subsection{Universal Adversarial Attack with Pre-Collected Input at the Receiver}
Here, we first relax the assumption that the adversary knows the exact input of the classifier. The adversary in the previous attacks generates an input-dependent perturbation, i.e., $\boldsymbol{\delta}$ is designed given the exact input $\boldsymbol{r}_{t}$. This requires the adversary to always know the input of the classifier, which is not a practical assumption to make due to synchronization issues. Thus, it is more practical to design an input-independent UAP. 

We propose an algorithm to design the UAP using PCA. We assume that the adversary collects some arbitrary set of inputs $\{\boldsymbol{r}^{(1)}_{t}, \boldsymbol{r}^{(2)}_{t},\cdot\cdot\cdot, \boldsymbol{r}^{(N)}_{t}\}$ and associated labels. The adversary generates perturbations $\{\boldsymbol{\delta}^{(1)}, \boldsymbol{\delta}^{(2)},\cdot\cdot\cdot,  \boldsymbol{\delta}^{(N)}\}$ with respect to the obtained arbitrary set of inputs and the exact channel information using schemes from the previous sections. To reflect the common characteristics of $\{\boldsymbol{\delta}^{(1)}, \boldsymbol{\delta}^{(2)},\cdot\cdot\cdot,  \boldsymbol{\delta}^{(N)}\}$ in the UAP, we stack these perturbations into a matrix and perform PCA to find the first component of the matrix with the largest eigenvalue. Hence, we use the direction of the first principal component as the direction of UAP for channel $\mathbf{H}_{ar}$. The algorithm for the UAP with $N$ pre-collected input data is similar to Algorithm \ref{limitch}. The difference is that there are $N$ pre-collected data inputs instead of $N$ realizations of the channel.

\subsection{Universal Adversarial Attack with Limited Channel Information}
Now, we further relax the assumption that the adversary knows the exact channel between the adversary and the receiver, and assume that the adversary only knows the distribution of this channel. To design the UAP knowing the distribution of the channel, we first generate random realizations of the channel $\{\mathbf{H}_{ar}^{(1)}, \mathbf{H}_{ar}^{(2)},\cdot\cdot\cdot, \mathbf{H}_{ar}^{(N)}\}$ from the distribution. Then we generate $\boldsymbol{\delta}^{(n)}$ using $\boldsymbol{r}_{t}^{(n)}$ and $\mathbf{H}_{ar}^{(n)}$ instead of using the real channel $\mathbf{H}_{ar}$ and real input $\boldsymbol{r}_{t}$. Again, we use PCA to find the first component of the matrix and use it as our direction of UAP. The algorithm for UAP with limited channel information is analogous to Algorithm \ref{limitch} except that we have pre-collected input data as opposed to real input data in Algorithm \ref{limitch}.

\subsection{Black-box Universal Adversarial Attack}
The last assumption that we will relax is the information about the classifier at the receiver. To relax this assumption, we use the well-known transferability property of adversarial attacks \cite{Transfer1}. This property states that the adversarial attack crafted to fool a specific DNN can also fool other DNNs with different architectures, with high probability. Therefore, the adversary generates UAPs using a substitute DNN and uses them to fool the actual DNN at the receiver.

\section{Simulation Results} \label{sec:SimResults}
We compare the performance of attacks proposed in this paper and another attack from \cite{Larsson2}.
We use VT-CNN2 classifier used in \cite{Oshea1} and \cite{Larsson2}, and train it with GNU radio ML dataset RML2016.10a \cite{dataset1}. The dataset contains 220,000 samples. Each sample corresponds to one specific modulation scheme at a specific signal-to-noise ratio (SNR). There are 11 modulations: BPSK, QPSK, 8PSK, QAM16, QAM64, CPFSK, GFSK, PAM4, WBFM, AM-SSB and AM-DSB. Also, we follow the same setup of \cite{Oshea1}, using Keras with TensorFlow backend, where the modulation classifier at the receiver estimates the modulation after receiving 128 I/Q (in-phase/quadrature) channel symbols. We assume that the channel between the adversary and the receiver is Rayleigh fading with path-loss and shadowing, i.e., $\boldsymbol{h}_{ar} = K(\frac{d_{0}}{d})^{\gamma}\psi \boldsymbol{h}_{ray}$ where $ K = 1, d_{0}=1, d=10, \gamma = 2.7, \psi \sim\ $Lognormal$(0,8)$ and $\boldsymbol{h}_{ray} \sim \mbox{Rayleigh}(0,1) $.

\begin{figure}[t]
	\centerline{\includegraphics[width=0.9\linewidth]{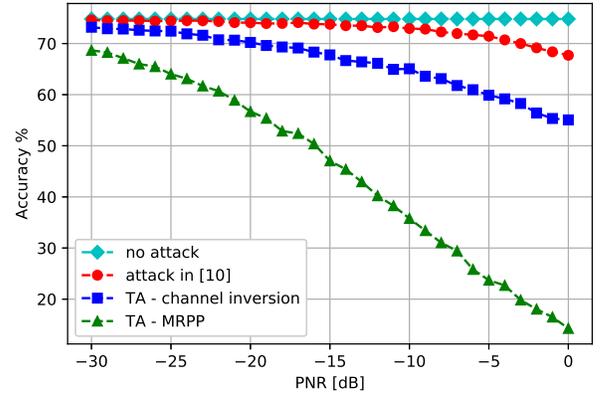}}
	\caption{Classifier accuracy with and without considering wireless channel when SNR = 10 dB.}
	\label{plot1}
\end{figure}

\begin{figure}[t]
	\centerline{\includegraphics[width=0.9\linewidth]{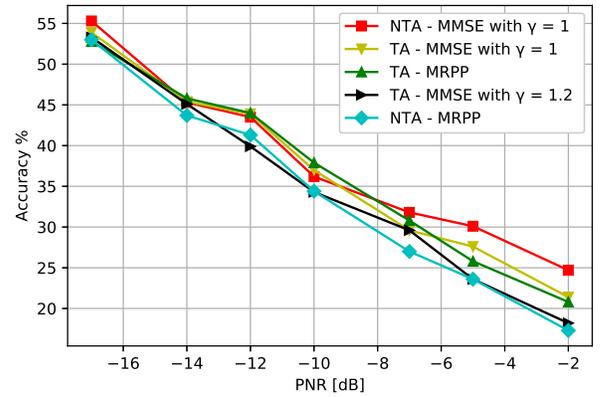}}
	\caption{Classifier accuracy under different white-box attack schemes when SNR = 10 dB.}
	\label{plot2}
\end{figure}
 Here, we use perturbation-to-noise ratio (PNR) metric from \cite{Larsson2} that shows the relative perturbation power with respect to the noise and measure how the increase in the PNR affects the accuracy of the classifier. Note that as the PNR increases, it is more likely to be detected by the receiver. In the figures, we denote targeted attack by TA and non-targeted attack NTA. 
 
 Fig. 1 presents the accuracy of the classifier versus PNR under the proposed targeted white-box adversarial attacks with exact channel information and the adversarial attack studied in \cite{Larsson2}. As expected, the white-box attack in \cite{Larsson2} considering only the AWGN channel has poor performance that is close to no attack case in low PNR region. The reason is that the wireless channel changes the phase and the magnitude of the perturbation at the receiver. Further, we see that the targeted channel inversion attack does not perform well compared to the targeted MRPP attack, indicating the importance of the received power at the classifier. 

The performance of the proposed white-box attacks is compared in Fig. 2. As discussed in Section IV.B, $\lambda$ can be optimized by using a line search method and it can be seen that the targeted MMSE attack performs better with $\lambda = 1.2$ compared to $\lambda = 1$. Furthermore, we observe that the non-targeted MRPP attack outperforms other attacks. This can be explained by the freedom of the direction that the non-targeted adversarial attack can take. For targeted attacks, we can only have 10 different directions since we have 11 modulation types, however, the non-targeted attacks do not have such restriction. Thus, it is more likely that the non-targeted attacks choose a better direction to enforce misclassification. Moreover, the computation complexity for non-targeted attacks is lower compared to the targeted attacks that involve iterations to reach the desired accuracy. 

In Fig. 3, we investigate the performance of the adversarial attacks with respect to different levels of information availability. First, we observe that the UAP with 40 pre-collected inputs, where the adversary knows the exact channel information, outperforms other attacks with limited information. This result shows the importance of the channel state information over the exact input data when crafting an adversarial attack. Note that the UAP with 40 pre-collected input data even outperforms the targeted channel inversion attack in the high PNR region, where the adversary knows not only the exact channel but also the exact input at the receiver. Furthermore, similar performance of the UAP with limited channel information and the black-box UAP shows transferability of adversarial attack, where for black-box UAP we use the same structure of the classifier but train it differently.

\begin{figure}[t]
	\centerline{\includegraphics[width=0.9\linewidth]{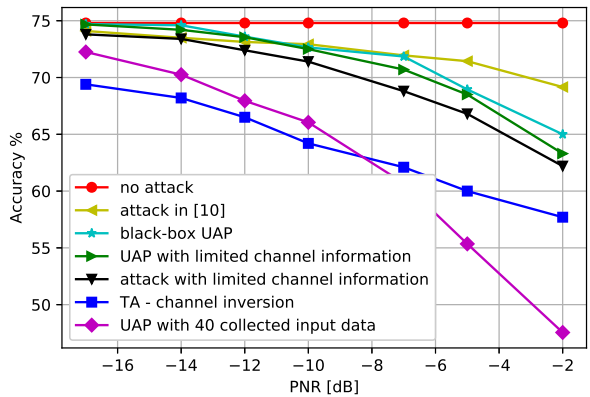}}
	\caption{Classifier accuracy using the UAP with different levels of information availability when SNR = 10 dB.}
	\label{plot3}
\end{figure}

\section{Conclusion} \label{sec:Conclusion}
We considered the wireless communication system, where DL algorithms are used to classify radio signals and showed that adversarial attacks against a modulation classifier are effective even when there are channel effects beyond the AWGN channel. Specifically, we considered both targeted attack and non-targeted attacks, and observed in the simulation results that DNNs used for modulation classification are vulnerable to these attacks. Furthermore, even with limited information, we show that the UAP can be generated to enforce misclassification at the receiver. 
%

%
%


\bibliographystyle{IEEEtran}
\bibliography{IEEEabrv,lib}

\end{document}